\documentclass[reprint,
amsmath,amssymb,
prl, 
]{revtex4-2}
\usepackage{graphicx}
\usepackage{dcolumn}
\usepackage{bm}
\usepackage[caption=false, font=large]{subfig}

\newcommand{\x}[0]{\mathbf{x}}
\newcommand{\dx}[0]{\mathrm{d}\x}
\DeclareMathOperator{\EX}{\mathbb{E}}

\graphicspath{{./images/}}

\begin{document}

	\title{Variationally Derived Intermediates for Correlated Free Energy Estimates between Intermediate States}
	
	\author{Martin Reinhardt}
	
	\author{Helmut Grubm\"uller}%
	\email{hgrubmu@gwdg.de}
	\affiliation{%
		Max Planck Institute for Biophysical Chemistry, Am Fassberg 11, 37077 G\"ottingen, Germany
	}%

	\date{\today}
	
	\begin{abstract}
	Free energy difference calculations based on atomistic simulations generally improve in accuracy when sampling from a sequence of intermediate equilibrium thermodynamic states that bridge the configuration space between two states of interest. For reasons of efficiency, usually the same samples are used to calculate the step-wise difference of such an intermediate to both adjacent intermediates. However, this procedure violates the assumption of uncorrelated estimates that is necessary to derive both the optimal sequence of intermediate states and the widely used Bennett acceptance ratio (BAR) estimator. In this work, via a variational approach, we derive the sequence of intermediate states and the corresponding estimator with minimal mean squared error that account for these correlations and assess its accuracy. 
	\end{abstract}
	
	\maketitle
	
	\section{Introduction}
	
    Free energy calculations are widely used to investigate physical and chemical processes \cite{Zuckerman2011, Jinnouchi2020, Sun2018, Ge2016}. Their accuracy is essential to biomedical applications such as computational drug development \cite{Christ2014, DeVivo2016, Cournia2017, Williams-Noonan2018} or material design \cite{Swinburne2018, Freitas2018, deKoning1999}. Amongst the most widely used methods based on simulations with atomistic Hamiltonians are alchemical equilibrium techniques, including the Free Energy Perturbation (FEP) \cite{Zwanzig1954} and Thermodynamic Integration (TI) \cite{Kirkwood1935} methods. These techniques determine the free energy difference between two states, representing, for example, two different ligands bound to a target, by sampling from intermediate states whose Hamiltonians are constructed from those of the end states. 
    
    The choice of these intermediates critically affects the accuracy of the free energy estimates \cite{Shenfeld2009, Zuckerman2002, Zuckerman2004} by determining which parts of the configuration space are sampled to which extent \cite{Pham2011}, thereby performing a function similar to importance sampling \cite{Gelman1998}. In addition, different estimators that determine the free energy differences between these intermediates and the end states have been developed, most prominently the Zwanzig formula \cite{Zwanzig1954} for FEP, the Bennett Acceptance Ratio method (BAR) \cite{Bennett1976}, and multistate BAR (MBAR) \cite{Shirts2008}. 

    We have recently derived \cite{Reinhardt2020} the sequence of discrete intermediate states that yields, for finite sampling, the lowest mean squared error (MSE) of the free energy estimates with respect to the exact value. Notably, minimizing the MSE accounts not only for the variance, but also for possible bias. The result differs from the most common scheme, which linearly interpolates between the end states Hamiltonians $H_1(\x)$  and $H_N(\x)$, respectively, along a path variable $\lambda$,
    \begin{align}
    H_s(\x) = (1-\lambda)H_1(\x, \lambda) + \lambda H_N(\x, \lambda), \;\; \lambda \in [0, 1] 
    \label{eq:lin_interpolation}
    \end{align} 
    where $\x\in {\rm I\!R}^{3M}$ denotes the coordinate vector of all $M$ particles in the system. Here, the additional $\lambda$ argument of the end states Hamiltonians indicates the commmon use of soft-core potentials \cite{Beutler1994, Zacharias1994, Steinbrecher2007} to avoid divergences for vanishing particle. Other approaches involve the interpolation of exponentially weighted Hamiltonians of the end states, such as Enveloping Distribution Sampling ~\cite{Christ2007} or the Minimum Variance path \cite{Blondel2004, Pham2012} for TI. 
    
    In contrast, the variationally derived intermediates (VI) turn out to be coupled and thus determined through a system of equations \cite{Reinhardt2020}. For the setup shown in Fig.~\ref{fig:setup}(a), where all states are labeled by integers $s$ with $1\leq s \leq N$, sampling is conducted in the intermediates with \textit{even} numbered $s$, governed by the optimal Hamiltonian
    \begin{align}
    \begin{split}
    &H_{s}(\x)\\
    &= -\frac{1}{2}\ln[e^{-2H_{s-1}(\x)}\cdot r_{s-1,s}^{-2} +e^{-2H_{s+1}(\x)}\cdot r_{s+1,s}^{-2}] \,.
     \end{split}
    \label{eq:soe_even}
    \end{align}	
    where $r_{s,t} = Z_s \slash Z_t$ denotes the ratio of the configurational partition sums of states $s$ and $t$. Virtual intermediates, i.e., the ones without sampling, are labeled with \textit{odd} $s$ with $2<s<N-1$ and indicated by the dashed lines in Fig.~\ref{fig:setup}(a). For these,
    \begin{equation}
    H_s(\x) = \ln[e^{H_{s-1}(\x)}\cdot r_{s-1,s} +e^{H_{s+1}(\x)}\cdot r_{s+1,s}] \,.
    \label{eq:soe_odd}
    \end{equation}
    Due to the dependence on the ratios of the partition sums, i.e., the desired quantity, the set of equations has to be solved iteratively. 
    The variational MSE minimization has been conducted based on the Zwanzig formula \cite{Zwanzig1954}
    \begin{align}
    \Delta G_{s,s+1}= -\ln\langle e^{-[H_{s+1}(\x) - H_s(\x)]}\rangle_s\,
    \label{eq:zwanzig}
    \end{align}
    being used to calculate the difference between two adjacent states, as indicated by the arrows in Fig.~\ref{fig:setup}. Furthermore, using the virtual target states described by Eq.~(\ref{eq:soe_even}) is equivalent to using BAR directly between two sampling states \cite{Lu2003, Reinhardt2020}, and, therefore, Eq.~\ref{eq:soe_corr_odd} also describes the optimal intermediates for BAR.  
    
    However, for BAR and VI to be optimal for multiple states, the free energy estimates to the states above and below an intermediate in the sequence have to be based on separate, uncorrelated sample points \cite{Reinhardt2020}, as illustrated by the separate yellow points in Fig.~\ref{fig:setup}(a) that we refer to as the regular FEP setup. Yet, it would be twice as efficient to use the same sample points in both directions, as illustrated by Fig.~\ref{fig:setup}(b), and as generally done in practice. However, this introduces correlations between the estimates to both adjacent intermediates, thereby violating the assumptions underlying the derivation of Eqs.~ (\ref{eq:soe_even}) and (\ref{eq:soe_odd}). Therefore, in this case the above variational intermediates are not optimal anymore. Due to these correlations, we refer to the Fig.~\ref{fig:setup}(b) as the correlated FEP (cFEP) setup. 
    
    Here, we derive the minimal MSE sequence of intermediate states for and the corresponding estimators for cFEP that take these correlations properly into account. As will be shown below, what might seem as a minor technical twist, markedly changes the shape of the optimal intermediates and considerably improves the accuracy of the obtained free energy estimates. 
   
    \section{Theory}
    
\begin{figure}[t!]
		\centering
		\includegraphics[width=0.8\linewidth]{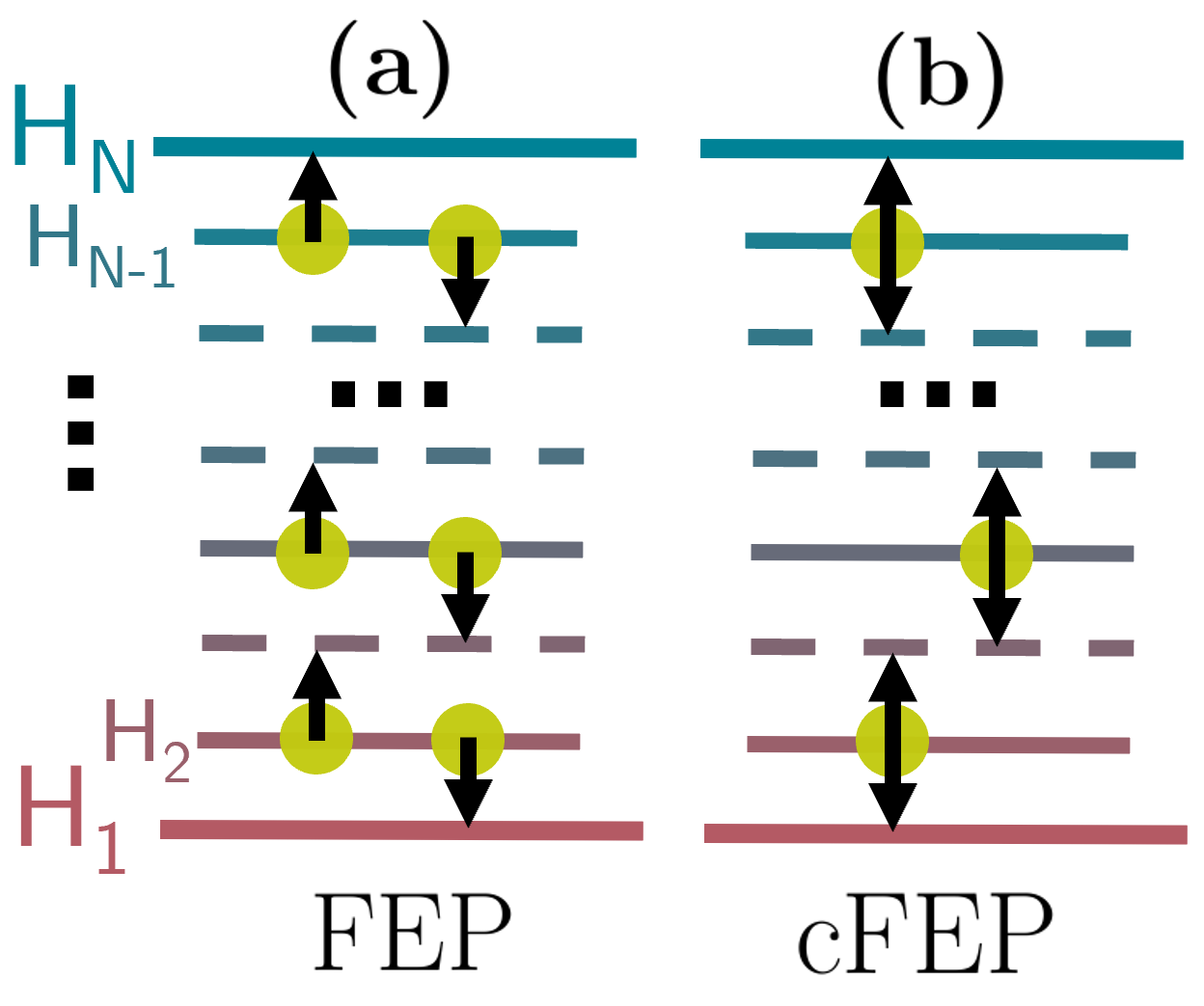}
		\caption{Two schemes of free energy calculation. The arrows indicate the Zwanzig formula is used to evaluate the free energy difference to the adjacent state based on sample sets represented through yellow dots. The dashed lines represent virtual intermediate states that no sampling is conducted in. (a) Separate and uncorrelated sample set are used to calculate the free energy difference of the respective intermediate to the state above and below (b) The same sample set is used for this purpose. }
		\label{fig:setup}
	\end{figure}

	For the cFEP scheme shown in Fig.~\ref{fig:setup}(b), using $N$ states, we aim to derive the sequence of intermediate Hamiltonians $H_2(\x)\ldots H_{N-1}(\x)$ that optimizes the MSE 
	\begin{equation}
	\mathrm{MSE}\left(\Delta G^{(n)}\right) = \EX\left[ \left(\Delta G - \Delta G^{(n)} \right)^2\right]
	\label{eq:acc_simple}
	\end{equation}
	along similar lines as before \cite{Reinhardt2020}. Here, $\Delta G^{(n)}_{1,N}$ denotes the free energy estimate based on a finite number of sample points $n$, and $\Delta G_{1,N}$ the exact difference between the end states 1 and $N$. 
	
	The cFEP variant in Fig.~\ref{fig:setup}(b) only uses sampling in the intermediate states. Setups that, in addition, involve sampling in the end states, can also be treated with the formalism below. However, firstly, as we have tested,  the accuracy for a given computational effort does not increase in this case. Secondly, mixing two different types of sample points (the ones used to evaluate $\Delta H$ to only one adjacent state vs. to both adjacent states) further complicates the analysis. 
	
	For cFEP, the estimated difference is
	\begin{equation}
	\Delta G^{(n)} = \sum_{\substack{s=2 \\ s \; \text{even}}}^{N-2} \left(\Delta G_{s \rightarrow s+1}^{(n)}  - \Delta G_{s \rightarrow s-1}^{(n)}\right) \,.
	\label{eq:acc_multiple}
	\end{equation} 
	As in Fig.~\ref{fig:setup}(b), the arrows point from sampling to target states, i.e., either the end states or the virtual intermediates. 
	Assuming for each sample state $s$ a set of $n$ independent sample points $\{\x_i\}$, drawn from ${p_s(\x)=e^{-H_s(\x)}\slash Z_s}$, 
	with partition function $Z_s$,
    Eq.~(\ref{eq:acc_multiple}) reads
	\begin{equation}
	\begin{split}
	&\mathrm{MSE}\left(\Delta G_{1,N}^{(n)}\right) =\:  \\
	 &\left(\Delta G_{1,N}\right)^2  + \sum_{\substack{s=2 \\ s \; \text{even}}}^{N-2}\EX\left[  \left(\Delta G_{s \rightarrow s+1}^{(n)} \right)^2 + \left(\Delta G_{s \rightarrow s-1}^{(n)} \right)^2\right] \\
	& - 2\Delta G_{1,N}\left(\sum_{\substack{s=2 \\ s \; \text{even}}}^{N-2}\left(\EX\left[\Delta G_{s \rightarrow s+1}^{(n)}\right] -\EX\left[\Delta G_{s \rightarrow s-1}^{(n)}\right]\right)\right)\\
	& - \sum_{\substack{s=2 \\ s \; \text{even}}}^{N-2}\;\sum_{\substack{t=2 \\ t \; \text{even}}}^{N-2}\EX\left[  2\,\Delta G_{s \rightarrow s+1}^{(n)}\,\Delta G_{t \rightarrow t-1}^{(n)}\right]\,.
	\end{split}
	\label{eq:msd_expanded}
	\end{equation}
	The first two lines of Eq.~(\ref{eq:msd_expanded}) have already been processed in Ref.~\citenum{Reinhardt2020}, but the last term differs. Previously, as in the regular FEP scheme in Fig.~\ref{fig:setup}(a), these last expectation values were originally derived from independent sample sets and were, therefore, uncorrelated. In the present context of cFEP, however, these estimates are correlated. Therefore, the term needs to be split in two sums, distinguishing between the pairs with samples from the same state and the ones from different states,
	\begin{equation}
	\begin{split}
	&\sum_{\substack{s=2 \\ s \; \text{even}}}^{N-2}\;\sum_{\substack{t=2 \\ t \; \text{even}}}^{N-2}\EX\left[  2\,\Delta G_{s \rightarrow s+1}^{(n)}\,\Delta G_{t \rightarrow t-1}^{(n)}\right] \\
	= & \;\;\;\;\,2\sum_{\substack{s=2 \\ s \; \text{even}}}^{N-2}\EX\left[ \Delta G_{s \rightarrow s+1}^{(n)}\,\Delta G_{s \rightarrow s-1}^{(n)}\right] \\
	&+2\sum_{\substack{s=2 \\ s \; \text{even}}}^{N-2}\;\sum_{\substack{t=2 \\ t \; \text{even} \\ t \neq s}}^{N-2}\EX\left[\Delta G_{s \rightarrow s+1}^{(n)}\right] \EX\left[\Delta G_{t \rightarrow t-1}^{(n)}\right] \,,
	\end{split}
	\label{eq:tmp}
	\end{equation}
	where the expectation value of the product between the two estimates based on different sample sets has been separated, as these are uncorrelated. 
	
	As we are only interested in the intermediates that optimize the MSE, and not in the absolute value of the MSE, we focus on the terms that will not drop out in the optimization below. 
	
	Continuing with the expression inside the sum of the first term on the right hand side of Eq.~\ref{eq:tmp},
	\begin{align}
	&\EX\left[ \Delta G_{s \rightarrow s+1}^{(n)}\,\Delta G_{s \rightarrow s-1}^{(n)}\right] \\
	\begin{split}
	=&- \int p_s(\x_1)\dx_1 ... \int p_s(\x_n)\dx_n\\
	& \ln \left[ \frac{1}{n} \sum_{i=1}^{n} e^{-(H_{s+1}(\x_i) - H_s(\x_i))}\right]\\
	&\ln \left[ \frac{1}{n} \sum_{i=1}^{n} e^{-(H_{s-1}(\x_i) - H_s(\x_i))}\right] \, .
	\end{split}
	\label{eq:correlated_expectation}
	\end{align}
	As in the derivation of Ref.~\citenum{Reinhardt2020}, the Hamiltonians are now shifted by a constant offset $C_s$, i.e., ${H_s'(\x) = H_s(\x) - C_s}$. This offset will cancel out for a given shape of an intermediate when calculating the accumulated free energy difference in Eq.~\ref{eq:acc_multiple}. However, as the intermediate states will turn out to be coupled, these offsets do influence the shape of these intermediates. The offsets can now be chosen such that the terms inside the logarithms of Eq.~(\ref{eq:correlated_expectation}) are close to one. In this case, $	{\EX\left[ \Delta G^{(n)}_{s' \rightarrow (s+1)'} \right]}  = { \Delta G_{s',(s+1)'} } $ \cite{Reinhardt2020}, and, therefore, 
	the two linear terms arising from Eq.~(\ref{eq:correlated_expectation}) can be expressed in terms of the exact free energy differences.

	Next, the product of the two sums in Eq.~\ref{eq:correlated_expectation} is split into terms based on the same and different sample points, respectively,
	\begin{align}
	&\EX\left[ \Delta G_{s' \rightarrow (s+1)'}^{(n)}\,\Delta G_{s' \rightarrow (s-1)'}^{(n)}\right] \\
	\begin{split}
	=&- \frac{1}{n^2}\int p_s(\x_1)\dx_1 ... \int p_s(\x_n)\dx_n\\
	&\left[\left(\sum_{i=1}^{n} e^{-(H'_{s+1}(\x_i) - H'_s(\x_i))}\right)
	\left(\sum_{\substack{j=1 \\ j \neq i}}^{n} e^{-(H'_{s-1}(\x_j) - H'_s(\x_j))}\right)\right.\\
	&\left.\;\;\; + \sum_{i=1}^{n} e^{-H'_{s+1}(\x_i) - H'_{s-1}(\x_i)+2H'_s(\x_i)}\;\;\right] \\
	&+ f_{s'}(\Delta G_{s'\rightarrow (s-1)'}, \Delta G_{s'\rightarrow (s+1)'}) \, ,
	\end{split}
	\label{eq:tmp2}
	\end{align}
	where the terms that can be expressed solely based on (constant) free energy differences are summarized by the term $f_s$. Again, the first two terms of Eq.~(\ref{eq:tmp2}) can be expressed in terms of the free energy differences between $s$ and $s+1$ as well as between $s$ and $s-1$, respectively. 
	
	Collecting all terms arising from Eq.~(\ref{eq:msd_expanded})
\begin{align}
\begin{split}
&\mathrm{MSE}\left(\Delta G_{1,N}^{(n)}\right)\\ 
=&  \sum_{\substack{s=2 \\ s \; \text{odd}}}^{N-2} \frac{1}{n}\Big(\int p_s(\x)\,\dx\, e^{-2(H'_{s+1}(\x)- H_s'(\x))}  \\ 
& + \int p_{s+2}(\x)\,\dx\, e^{-2(H'_{s+1}(\x)- H'_{s+2}(\x))} \\ 
& + \int p_{s+1}(\x)\,\dx\, e^{-H'_{s+2}(\x)- H'_s(\x) + 2H_{s+1}(\x)} \\ 
& + g_{s'}(\Delta G_{s',(s+1)'}, \Delta G_{(s+2)',(s+1)'}, \Delta G_{1',N'})\; \Big) \,,
\end{split}
\label{eq:msd_squared_terms}
\end{align}
where the function $g_s'$ serves the same purpose as $f_s'$ and can be dropped in the optimization below.  
\begin{figure*}
	\subfloat[$N=3$ states]{		\includegraphics[width=0.49\textwidth]{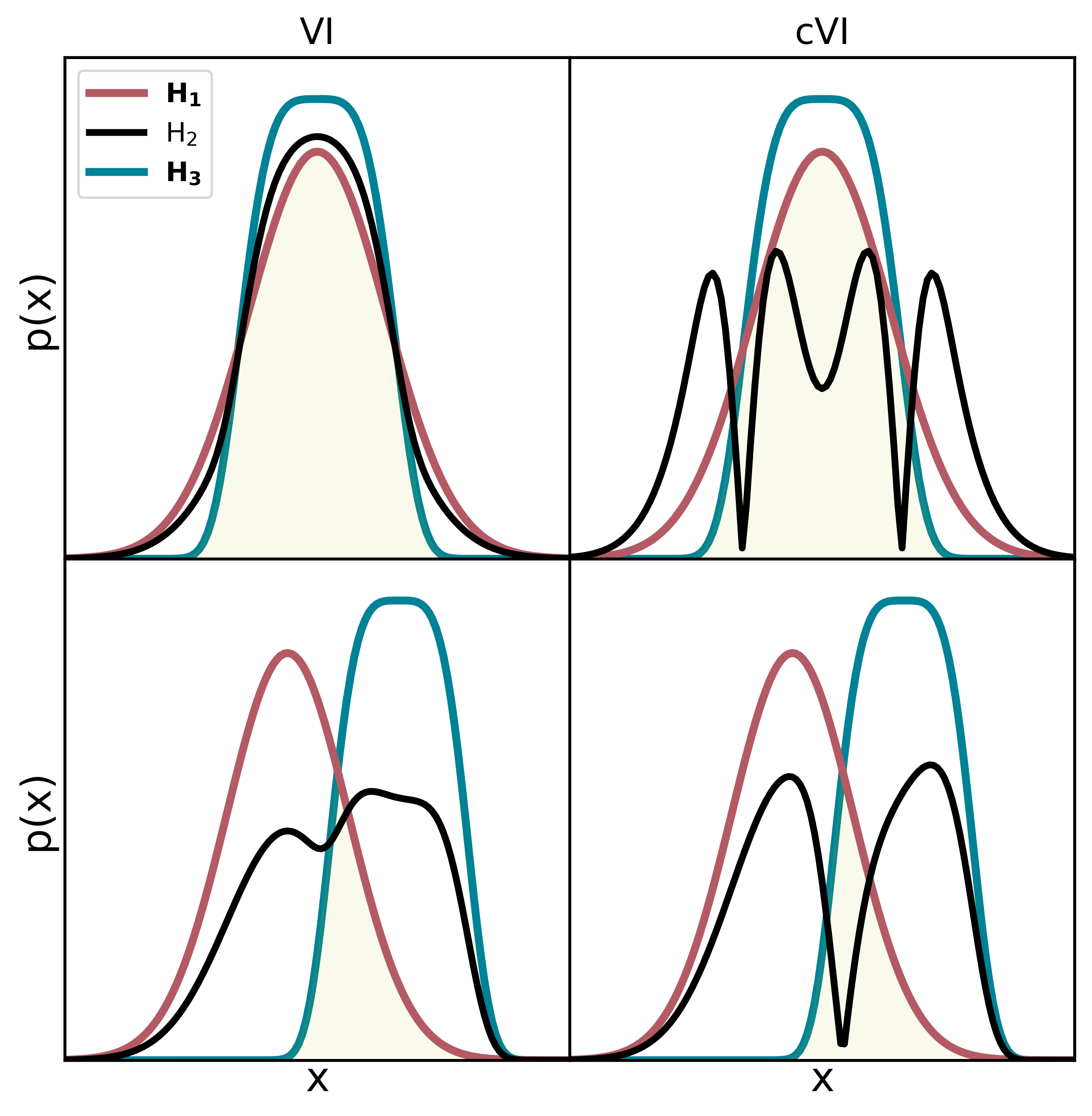}\label{fig:1}} 
	\subfloat[$N=7$ states]{		\includegraphics[width=0.49\textwidth]{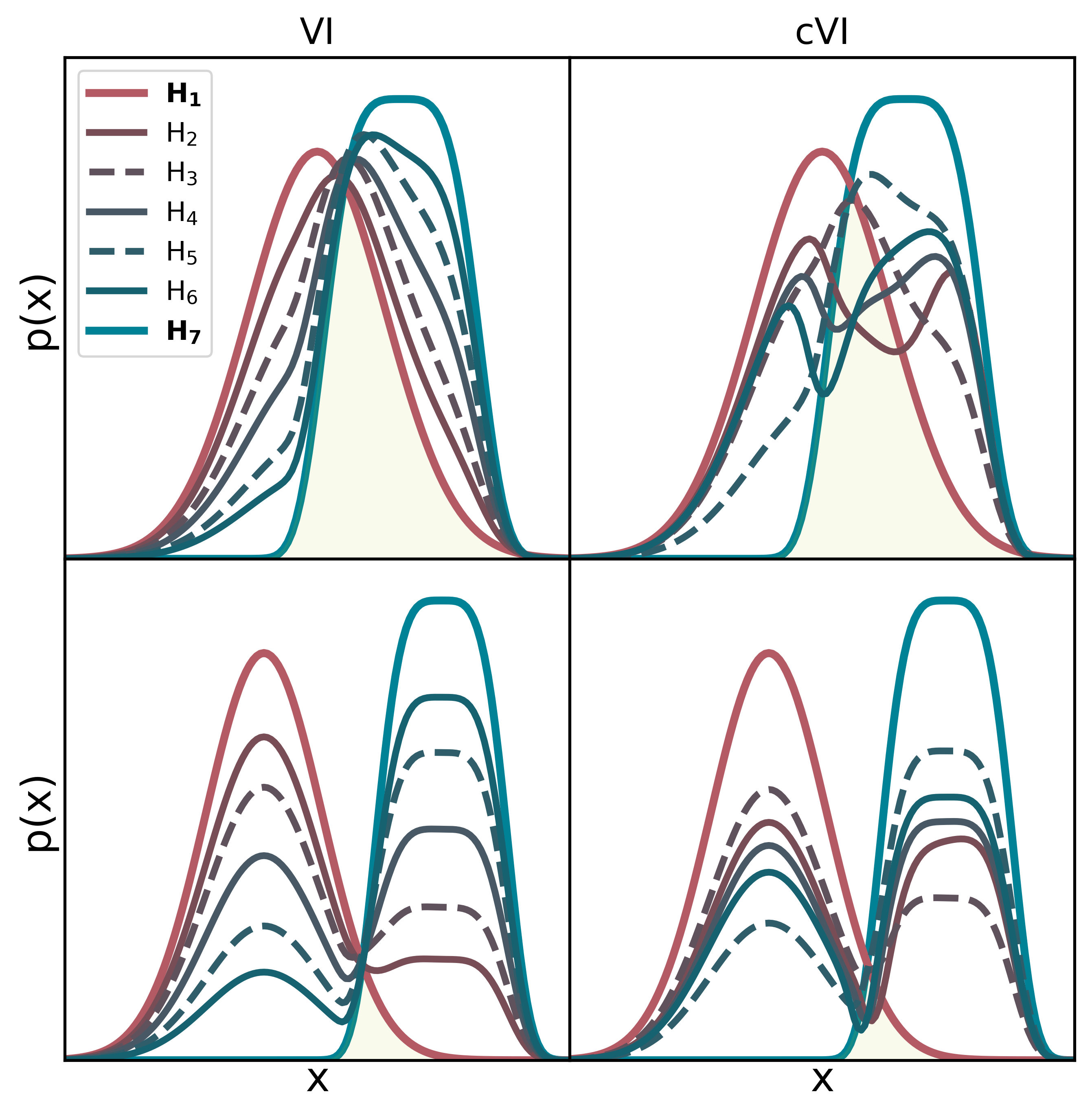}\label{fig:2}}
	\caption{Configuration space densities of VI (left column), and cVI (right column). The individual rows in (a) and (b) show different shifts in x-direction between the minima of the harmonic, $H_1(\x)$, and the quartic, $H_N(\x)$, potentials of the end states, thereby showing setups with different configuration space density overlap $K$ between the end states, indicated by the yellow area. Sampling is conducted in the even numbered intermediates. The dashed lines in (b) indicate the (odd numbered) virtual intermediate target states that no sampling is conducted in.}\label{fig:phase_space}
\end{figure*}

The condition of small $\Delta G^{(n)}_{s'\rightarrow (s+1)'}$ is fulfilled by setting $C_s = -\ln Z_s$.  By variation of the MSE from Eq.~(\ref{eq:msd_squared_terms}), 
\begin{equation}
\frac{\partial}{\partial H_s(\x)}\left(\mathrm{MSE}\left(\Delta G_{1,N}^{(n)}\right) + \nu \int (e^{-H_s(\x)} - Z_s) d\x\right) \overset{!}{=} 0 \, ,
\end{equation}
where $\nu$ is a Lagrange multiplier, the optimal sequence of Hamiltonians is obtained. For $s$ \textit{even}, we obtain
\begin{align}
	\begin{split}
		H_s(\x) = -\frac{1}{2}\ln\left(e^{-2H_{s-1}(\x)}r_{s-1,s}^{-2} + e^{-2H_{s+1}(\x)}r_{s+1,s}^{-2} \right.\\
		\left. -2 e^{-H_{s-1}(\x) -H_{s+1}(\x)}r_{s-1,s}^{-1}r_{s+1,s}^{-1}\right)
	\end{split}
	\label{eq:soe_corr_even}
\end{align}	
For $s$ \textit{odd} and $2<s<N-1$: 	
\begin{align}
	\begin{split}
		H_s(\x) = \; \ln &\left(e^{H_{s-1}(\x)}r_{s-1,s} + e^{H_{s+1}(\x)}r_{s-1,s}\right) \\
		 - \ln&\Big( e^{-H_{s-2}(\x)+H_{s-1}(\x)}r_{s-1,s-2}\\
		 & + e^{-H_{s+2}(\x)+H_{s+1}(\x)} r_{s+1,s+2}\Big)
	\end{split}
	\label{eq:soe_corr_odd}
\end{align}
where, as in Eqs.~(\ref{eq:soe_even}) and (\ref{eq:soe_odd}), the ratios $r_{s,t}$ of the partition sums between states $s$ and $t$ have to be determined iteratively. The above sequence, Eqs.~(\ref{eq:soe_corr_even}) and (\ref{eq:soe_corr_odd}), that we refer to as the correlated Variational Intermediates (cVI), yield the minimal MSE estimates for cFEP. 

Figure~\ref{fig:phase_space} shows the resulting configuration space densities of the above intermediates for the example of a start state with a harmonic Hamiltonian, $H_1(\x) = \frac{1}{2}x^2$, and an end state with a quartic one, $H_N(\x)=(x-x_0)^4$. 
Panel~(a) shows the VI that are optimal for the regular FEP scheme in Fig.~\ref{fig:setup}(a). Panel~(b) shows the cVI, optimal for cFEP.

The yellow areas in Fig.~\ref{fig:phase_space}, Eq.~(\ref{eq:overlap}), provide a simple measure of the configuration space density overlap $K$ between the end states 1 and $N$, 
\begin{equation}
K = \int_{-\infty}^{+\infty}d\x\:\mathrm{min}(p_A(\x), p_B(\x)) \, ,
\label{eq:overlap}
\end{equation}
Here, $K=0$ indicates two separate distributions without any overlap, and $K=1$ full overlap, i.e., identical configuration space densities.

The two rows in Fig.~\ref{fig:phase_space}(a) and (b) depict the result for two different values of $x_0$, and correspondingly, varying $K$.  

As can be inferred from Eq.~(\ref{eq:soe_corr_even}), for $N=3$, $H_2(\x)$ diverges at the points where ${p_1(\x)=p_3(\x)}$, and therefore, $p_2(\x) = 0$ at these points, as can also be seen for the intermediate sampling state shown in  Fig.~\ref{fig:phase_space}(a). More generally, $H_2(\x)$ of cVI ``directs'' sampling away from the overlap regions and towards the ones that are only relevant for one, but not both end states. For instance, the tails of the start state in the upper row of (a) are sampled more for cVI than for VI. For larger horizontal shifts of $x_0$, i.e., low values of $K$, the two variants become increasingly similar, as the additional term in Eq.~(\ref{eq:soe_corr_even}) with respect to Eq.~(\ref{eq:soe_even}) becomes smaller compared to the first term. 

For ${N=7}$ states, Fig.~\ref{fig:phase_space}(b) shows the converged resulting configuration space densities. The case of ${x_0=0}$, as shown in (a), was omitted in (b) as the visualization is more difficult in this case due to the higher number of states. In (b), the additional changes from VI to cVI become more complex. As in (a), the sampling states have smaller densities $p(\x)$ in the overlap regions of the end states, but, in contrast to (a), still differ between VI and cVI for smaller values of overlap $K$.  The reason is that while the overlap between the end states vanishes with decreasing $K$, an overlap between adjacent intermediate states remains that affects the shape of the intermediates. 
Note that the divergences mentioned above introduce instabilities in solving the system of Eqs.~(\ref{eq:soe_corr_even}) and (\ref{eq:soe_corr_odd}). Hence, for $N>3$ the factor~2 of the additional term in the logarithm Eq.~(\ref{eq:soe_corr_even}) has been replaced by a factor $\kappa$ that was set to slightly below 2 (${\kappa = 1.95}$) in case of Fig.~\ref{fig:phase_space}(b). See Appendix A for details.

\subsection{cBAR Estimator}

As mentioned above, using the Zwanzig formula \cite{Zwanzig1954} to evaluate the free energy difference between two sampling states with respect to the virtual intermediate, Eq.~(\ref{eq:soe_odd}), of VI is equivalent to BAR \cite{Reinhardt2020, Lu2003}. Correspondingly, the virtual intermediate defined by Eq.~(\ref{eq:soe_corr_odd}) of cVI also corresponds to an estimator, that is optimal for the sampling states of cFEP and that we will refer to as correlated BAR (cBAR).  

To derive cBAR, we use the relation between the two approaches. Determining the free energy difference between two sampling states labeled $s-1$ and $s+1$ by using the virtual intermediate $s$ to evaluate the difference between the adjacent states yields
\begin{align}
\Delta G^{(n)}_{s-1, s+1} = -\ln \frac{\langle e^{-(H_s(\x) - H_{s+1}(\x))}\rangle_{s+1}}{\langle e^{-(H_s(\x) - H_{s-1}(\x))}\rangle_{s-1}} \, .
\label{eq:fep_approach}
\end{align}
Using the approach of Bennett \cite{Bennett1976} instead,
\begin{align}
	&\Delta G^{(n)}_{s-1, s+1} \nonumber\\
	= &\ln \frac{\langle w(H_{s-1}(\x), H_{s+1}(\x)) e^{-H_{s-1}(\x)}\rangle_{s+1}}{\langle w(H_{s-1}(\x), H_{s+1}(\x)) e^{-H_{s+1}(\x)}\rangle_{s-1}} \, .
	\label{eq:bar_weighting}
\end{align}
where $w(H_{s-1}(\x), H_{s+1}(\x))$ is a weighting function. From Eqs.~(\ref{eq:zwanzig_two_step}) and (\ref{eq:bar_weighting}) follows that the two approaches are equivalent if the weighting function relates to the Hamiltonian of the virtual intermediate state through 
\begin{align}
	w(H_{s-1}(\x), H_{s+1}(\x)) = e^{-H_s(\x) + H_{s-1}(\x) + H_{s+1}(\x)} \,.
	\label{eq:equivalence}
\end{align}
 Therefore, any Hamiltonian of a virtual intermediate state corresponds to a weighting function. Bennett optimized the weighting function with respect to the variance yielding the famous BAR result
\begin{align}
	\Delta G^{(n)}_{s-1, s+1} - C= \ln \frac{\langle f(H_{s-1}(\x) - H_{s+1}(\x) - C)\rangle_{s+1}}{\langle f(H_{s+1}(\x) - H_{s-1}(\x) + C)\rangle_{s-1}} \, ,
	\label{eq:zwanzig_two_step}
\end{align}
where $C\approx \Delta G_{s-1, s+1}$ has to be determined iteratively and $f(x)$ is the Fermi function. This result is equivalent to using the virtual intermediate of Eq.~(\ref{eq:soe_odd}) with Eq.~(\ref{eq:fep_approach}). Note that the relation of a virtual intermediate to BAR result had already been obtained by \citet{Lu2003}, albeit through a different formalism, and that using the hyperbolic secant function (Eq.~10, p. 2980), in their Overlap Sampling approach \cite{Lu2003, Lu2004a} is equivalent to Eq.~(\ref{eq:equivalence}).  

Next, for cFEP, using the Hamiltonian of the virtual intermediate from Eq.~(\ref{eq:soe_corr_odd}) in Eq.~(\ref{eq:equivalence}) yields the weighting function of cBAR,
\begin{align}
\begin{split}
&w\big(H_{s-2}(\x), H_{s-1}(\x), H_{s+1}(\x), H_{s+2}(\x), \\
&\;\;\;\;\; C_{s-2,s-1}, C_{s-1,s+1}, C_{s+1,s+2}\big) \\
= &\Big(e^{-H_{s-2}(\x) + H_{s-1}(\x) + C_{s-2, s-1}}\\ 
&\;\; e^{-H_{s+2}(\x) + H_{s+1}(\x) + C_{s+2, s+1}}\Big) \Big/ \\
&\left( e^{H_{s-1}(\x) - H_{s+1}(\x) - C_{s+1, s-1}} + 1 \right) \,,
\end{split}
\label{eq:alternative_bar}
\end{align}
where the MSE of the resulting estimates is minimal if all $C_{s, t} \approx \Delta G_{s,t}$.  A numerator of 1 in Eq.~\ref{eq:alternative_bar} would yield the original BAR result. 

Note that $H_{s-2}(\x)$, and $H_{s+2}(\x)$, are also virtual intermediates determined by Eq.~\ref{eq:soe_corr_odd}. As such, the result is a system of weighting functions, i.e., one for every pair of adjacent sampling states. The optimal estimate can, therefore, only be found by iteratively solving for the free energy estimates between all sampling states at once. In this regard, the procedure is similar to MBAR \cite{Shirts2008}.

\section{Test Simulations}		
	
To assess to what extent our new variational scheme improves accuracy, we consider the one-dimensional system with a harmonic and a quartic end state shown in Fig.~\ref{fig:phase_space}. Rejection sampling is used to obtain uncorrelated sample points. The free energy estimate, obtained from these finite sample sets, is compared to the exact free energy difference. The MSE, Eq.~(\ref{eq:acc_simple}), is then calculated by averaging over one million of such realizations. With this procedure, different combinations of overlap $K$, numbers of states $N$ and sample points $n$ are considered. 

We compare three variants. Firstly, using VI, Eqs.~(\ref{eq:soe_even}) and (\ref{eq:soe_odd}), with FEP, i.e., the scheme in Fig.~\ref{fig:setup}(a). Here, the estimates to both adjacent states are based on separate sample sets and, therefore, not correlated.
Secondly, also using VI, but now with cFEP, shown in Fig.~\ref{fig:setup}(b). In contrast to variant 1, these estimates are based on the same sample sets and, therefore, correlated. In order to keep the total computational effort constant, the number of sample points per set (i.e., per yellow point in Fig.~\ref{fig:setup}) is two times larger for cFEP than for FEP. Thirdly, using cVI, Eqs.~(\ref{eq:soe_corr_even}) and (\ref{eq:soe_corr_odd}), that accounts for these correlations, also with cFEP.

\begin{figure*}
	\subfloat{\raisebox{1ex}{\includegraphics[width=0.49\textwidth]{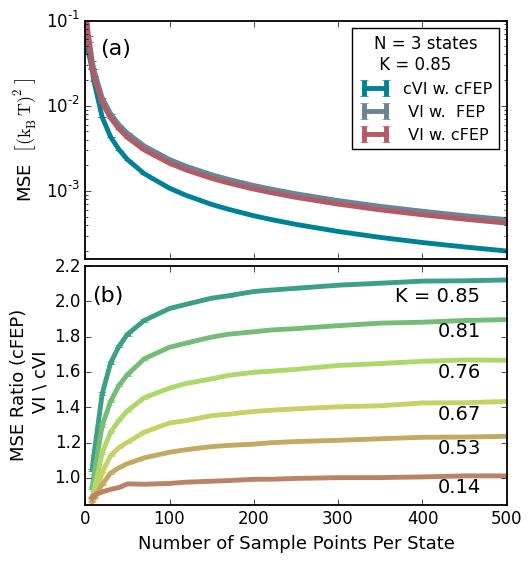}}} 
	\subfloat{	\includegraphics[width=0.49\textwidth]{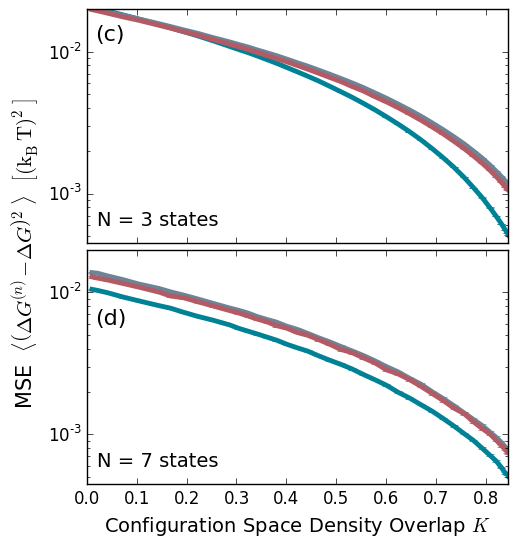}}
	\caption{Comparison of the accuracy of VI and cVI using the schemes of Fig.~\ref{fig:setup}. The accuracies were obtained from test simulations based on the setups shown in Fig.~\ref{fig:phase_space}. (a) Using $N=3$ states and comparing three variants of free energy calculations: Using cVI with cFEP (blue), VI with cFEP (red) and VI with FEP (grey). The MSEs of free energy calculations are shown for different number of sample points. (b) The ratio of the MSEs, and therefore, the improvement, of using cVI compared to VI for cFEP. The dark green line ($K=0.85$) corresponds to the ratio between the red and the blue line in (a). In addition, the results for different configuration space density overlaps $K$ between the end states are shown (green to orange). (c) Using $n=200$ sample points, the MSEs of the three variants from (a) are shown over the full range of $K$. (d) As in (c), but with $N=7$ states. The computational effort was kept constant by reducing the number of sample points per state. }
	\label{fig:results}
\end{figure*}

\section{Results}

For $N=3$ states, Fig.~\ref{fig:results}(a) shows the MSEs of the three variants for different numbers of sample points. Here, for the quartic end state, $x_0=0$, corresponding to $K=0.85$, was used. The corresponding configuration space densities of VI and cVI are shown in the upper row of Fig.~\ref{fig:phase_space}(a). 

As can be seen, cVI with cFEP, shown by the dark blue line, yields the best MSE for all numbers of sample points except very few ones. The other two variants, i.e., VI with FEP (grey line) and cFEP (red line) yield very similar MSEs. As such, the gain in information from evaluating the Hamiltonians to both adjacent states for all sample points yields only a very small improvement compared to using  separate sample sets for this purpose. 
	
In order to quantify the improvement of cVI compared to VI for cFEP, Fig.~\ref{fig:results}(b) shows the ratio of the MSEs of the two variants, again in relation to the number of sample points per set. The dark green curve ($K=0.85$), corresponds to the MSEs shown in (a) (i.e., the values of the red curve divided by the blue curve). The improvement in the MSE plateaus slightly above two for more than two hundred sample points per state. In addition, the improvements for setups with different overlap $K$ between the end states are shown (orange to light green). This improvement becomes smaller for smaller values of $K$, but the qualitative dependence on the number of sample points remains the same. 

For a constant number of sample points $n=200$ (and $n=100$ per set for VI with FEP, shown in grey), Fig.~\ref{fig:results}(c) shows how the MSEs of the three variants improve with increasing $K$. The MSEs converge at low $K$, which is in agreement with the observation from Fig.~\ref{fig:phase_space}(a) that the phase space densities of the intermediate state become more similar in this case. 

Figure~\ref{fig:results}(d) shows the MSEs for $N=7$ states. The corresponding configuration space densities for two different values of $K$ are shown in Fig.~\ref{fig:phase_space}(b). Here, VI with FEP and cFEP still yield similar MSEs, whereas cVI with cFEP, in contrast to $N=3$, now yields the best MSE for all $K$. The improvement to VI ranges from around 20~\%\ for low $K$, to around 50~\% for large $K$. This is in line with the observation from Fig.~\ref{fig:phase_space}(b) that the configuration space densities between VI and cVI become more similar but do not fully converge for a larger number of states in the limit of small $K$. 


Lastly, the cBAR estimator can be used with any choice of intermediate states for cFEP. To assess how much the cBAR estimator improves the accuracy of free energy estimates compared to BAR for cFEP, we conducted test simulations where the sampling states were chosen as in Eq.~(\ref{eq:lin_interpolation}), i.e., by linear interpolation between the Hamiltonians of the end states. Test simulations were conducted at varying values of $K$ and at ${N=5}$ and ${N=7}$. Evaluating the MSE, we found a statistically significant improvement, however, only in the range of ${1-2}$~\% (data therefore not shown here). The improvement was independent of $K$ and similar for both numbers of $N$. 

Considering that the MSEs of cVI and VI can improve up to an order of magnitude compared to the linear intermediates defined in Eq.~(\ref{eq:lin_interpolation}) (for a detailed comparison between VI and linear intermediates, see Ref.~\citenum{Reinhardt2020}), the large majority of improvements is not due to an improved estimator, but due to the way samples are generated.

\section{Discussion and Conclusion}	

In summary, we have derived a new variant of variational intermediates (cVI) that yield the optimal free energy estimate with minimal MSE when using the same sample points to evaluate the differences between the adjacent states above and below in the sequence (cFEP). This procedure is commonly used in free energy simulations, as it is computationally much cheaper to evaluate sample points at different Hamiltonians than to generate these. However, the resulting correlations between these estimates have not been considered yet. 

Our test simulations for a one-dimensional Hamiltonian show that cVI with cFEP yields an improved MSE compared to the optimal sequence (VI) with FEP, i.e., using different sample points for estimates to states above and below in the sequence. For $N=3$ states, the first variant improved the MSE by more than a factor of two for end states with high configuration space density overlap $K$, whereas at low $K$ the MSEs were similar. For $N=7$ states, the MSE improved between 20~\%\ (low $K$) and 50~\%\ (large $K$). 

Interestingly, due to the correlations mentioned above, using VI with FEP yields only slightly worse MSEs for all $K$ as using VI with cFEP, even though the latter involves twice as many evaluations of  Hamiltonians from adjacent states. Only for cVI,  thereby accounting for these correlations, the additional gain in information translates into a marked improvement of the MSE. 

Similar to most other theoretical analyses and derivations of free energy calculation methods, we also needed to assume that all sample points within each intermediate state are uncorrelated. If atomistic simulations are used for sampling, the resulting time-correlations reduce the number of essentially independent sample points. Unfortunately, for our one-dimensional systems, cVI increases barrier heights, thereby increasing correlation times. We have so far not tested our method on any complex biomolecular systems, so it is unclear if these barriers can be circumvented or what the expected increase in correlation times is. However, to avoid such correlations between sample points in atomistic simulations, usually only a small subset of all sample points is used to calculate free energy differences. Based on our findings and in contrast to common practice, we therefore recommend to use different subsets to evaluate the free differences to different adjacent states. 

The above derivation provides an example on how optimal intermediates and estimators with minimal MSE can be derived for different types of setups based on finite sampling that may help to incorporate a variety of assumptions and models into future theoretical approaches. 
	
\section{Appendix A: Avoiding numerical instabilities}
	
The divergence in Eq.~(\ref{eq:soe_corr_even}) at all $\x$ for which
\begin{align}
\begin{split}
&e^{-2H_{s-1}(\x)}r_{s-1,s}^{-2} + e^{-2H_{s+1}(\x)}r_{s+1,s}^{-2} \\
=\; &2 e^{-H_{s-1}(\x) -H_{s+1}(\x)}r_{s-1,s}^{-1}r_{s+1,s}^{-1}
\end{split}
\end{align}
causes numerical instabilities in solving the system of Eqs.~(\ref{eq:soe_corr_even}) and (\ref{eq:soe_corr_odd}). Replacing the factor 2 in Eq.~(\ref{eq:soe_corr_even}) in the logarithm with a factor $\kappa$, i.e., for $s$ even,
\begin{align}
\begin{split}
H_s(\x) = -\frac{1}{2}\ln\left(e^{-2H_{s-1}(\x)}r_{s-1,s}^{-2} + e^{-2H_{s+1}(\x)}r_{s+1,s}^{-2} \right.\\
\left. -\kappa e^{-H_{s-1}(\x) -H_{s+1}(\x)}r_{s-1,s}^{-1}r_{s+1,s}^{-1}\right)\, ,
\end{split}
\label{eq:soe_corr_even_kappa}
\end{align}
and setting, e.g., $\kappa = 1.95$, avoids these complications. As can be easily validated, the inside of the logarithm in Eq.~\ref{eq:soe_corr_even_kappa} is larger than zero for $0<\kappa<2$ for all $H_{s-1}(\x)$ and $H_{s+1}(\x)$.	
As shown for cVI in Fig.~\ref{fig:phase_space}(b), $\kappa<2$ prevents $p_s(\x)$ to go to zero at the crossing points of $p_{s-1}(\x)$ and $p_{s+1}(\x)$ of the neighboring states, but is still lowered at these points.

\bibliography{corr_deltaH}

\end{document}